\documentclass[reqno]{amsart}
\usepackage{amssymb,amsmath,amsthm}
\usepackage{eucal}
\usepackage{color}

\renewcommand{\theequation}{\arabic{equation}}
\begin{document}
\newtheorem{remark}{Remark}[section]

\def\upbrall{$\m@th\bracell$}
\def\undertilde#1{\mathop{\vtop{\ialign{##\crcr
    $\hfil\displaystyle{#1}\hfil$\crcr
     \noalign
     {\kern1.5pt\nointerlineskip}
     \upbrall\crcr\noalign{\kern1pt
   }}}}\limits}
\def\theequation{\arabic{section}.\arabic{equation}}
\newcommand{\ar}{\alpha}
\newcommand{\bb}{\beta}
\newcommand{\gm}{\gamma}
\newcommand{\Gm}{\Gamma}
\newcommand{\en}{\epsilon}
\newcommand{\dd}{\delta}
\newcommand{\sg}{\sigma}
\newcommand{\kp}{\kappa}
\newcommand{\ld}{\lambda}
\newcommand{\oa}{\omega}
\newcommand{\be}{\begin{equation}}
\newcommand{\ee}{\end{equation}}
\newcommand{\bea}{\begin{eqnarray}}
\newcommand{\eea}{\end{eqnarray}}
\newcommand{\bse}{\begin{subequations}}
\newcommand{\ese}{\end{subequations}}
\newcommand{\nn}{\nonumber}
\newcommand{\ol}{\overline}
\newcommand{\wt}{\widetilde}
\newcommand{\ut}{\undertilde}
\newcommand{\ip}{{i^\prime}}
\newcommand{\jp}{{j^\prime}}
\newcommand{\cn}{{\rm cn}}
\newcommand{\dn}{{\rm dn}}
\newcommand{\mbe}{{\boldsymbol e}}
\newcommand{\dint}{\int_\Gamma d\mu(\ell) }

\newcommand\Ptwo{\hbox{P}_{\scriptsize
       \hbox{II}}}
\newcommand\Pfour{\hbox{P}_{\scriptsize
       \hbox{IV}}}

\def\hypotilde#1#2{\vrule depth #1 pt width 0pt{\smash{{\mathop{#2}
\limits_{\displaystyle\widetilde{}}}}}}
\def\hypohat#1#2{\vrule depth #1 pt width 0pt{\smash{{\mathop{#2}
\limits_{\displaystyle\widehat{}}}}}}
\def\hypo#1#2{\vrule depth #1 pt width 0pt{\smash{{\mathop{#2}
\limits_{\displaystyle{}}}}}}

\newcommand{\pii}{P$_{{\rm\small II}}$}  
\newcommand{\pvi}{P$_{{\rm\small VI}}$}   

\newcommand{\pp}{\partial}
\newcommand{\hf}{\frac{1}{2}}
\newcommand{\ith}{$i^{\rm th}$\ }

 \def\hypotilde#1#2{\vrule depth #1 pt width 0pt{\smash{{\mathop{#2}
 \limits_{\displaystyle\widetilde{}}}}}}
 \def\hypohat#1#2{\vrule depth #1 pt width 0pt{\smash{{\mathop{#2}
 \limits_{\displaystyle\widehat{}}}}}}
 \def\hypo#1#2{\vrule depth #1 pt width 0pt{\smash{{\mathop{#2}
 \limits_{\displaystyle{}}}}}}

\newtheorem{theorem}{Theorem}[section]
\newtheorem{lemma}{Lemma}[section]
\newtheorem{cor}{Corollary}[section]
\newtheorem{prop}{Proposition}[section]
\newtheorem{definition}{Definition}[section]
\newtheorem{conj}{Conjecture}[section]
\newcommand{\bpr}{\begin{prop}}
\newcommand{\epr}{\end{prop}}

\title{\bf Direct ``delay'' reductions of the Toda hierarchy}
\author{N. Joshi}
\address{School of Mathematics and Statistics F07, The University of Sydney, NSW 2006
Australia}
\email{nalini@maths.usyd.edu.au}
\author{P.E. Spicer}
\address{Katholieke Universiteit Leuven, Departement Wiskunde, Celestijnenlaan 200B
B-3001 Leuven, Belgium}
\email{Paul.Spicer@wis.kuleuven.be}

\thanks{
The authors thank the Issac Newton Institute for Mathematical Sciences, where this paper was completed. We also wish to thank Pavlos Kassotakis for insightful conversations.  This research is supported by the
Australian Research Council Discovery Project Grant \#DP0664624.}

\begin{abstract}
We apply the direct method of obtaining reductions to the Toda hierarchy of equations.
The resulting equations form a hierarchy of ordinary differential difference equations, also known as delay-differential equations. Such a hierarchy appears to be the first of its kind in the literature.
All possible reductions, under certain assumptions, are obtained.
The Lax pair associated to this reduced hierarchy is obtained.
\end{abstract}
\maketitle

\section{Introduction}
\setcounter{equation}{0} The Toda equation and its hierarchy form important lattice models in physics.  Their applications range from the study of thermalization
in metals to the study of cellular neural networks and optical lattices. This has led to widespread interest in the study of their solutions. Reductions provide us with
one method of extending our knowledge of solutions. In this paper, we develop a new approach of deducing reductions that is based on the direct method.

The Toda equation is also important in the theory of the Painlev\'e equations \cite{Nou,Oka}. It appears when the B\"acklund transformations of the second and fourth Painlev\'e equations are iterated in the space of parameter values. Any reductions of the Toda equation therefore reflect the behaviour of the corresponding iterated Painlev\'e transcendents along invariant curves in the space coordinatized by the independent variable of the Painlev\'e equation and the space of its parameters.

Reductions based on the method of Lie symmetries \cite{LW} and the method of conditional symmetries \cite{LW1} are already known for the Toda equation, although
neither appears to have been applied to the Toda hierarchy. While in the case of PDEs, the results of the direct method are known to be included in the results of
the method of conditional symmetries, we note that no such relationship is known for differential-difference equations. By using the direct method, we find reductions
that lead to delay-differential equations in which the solutions are iterated along arbitrary curves.

The Toda hierarchy appears to have been described for the first time in the literature in 1983 \cite{UT}, although its associated recursion operator may have
been known earlier \cite{Dodd}.
The Toda hierarchy can be presented as
\be\label{eq:Toda}
\textrm{TL}_{r}(u,v) = \left(\begin{array}{c}
u_{t}-u(\ol{p}_{r+1}-p_{r+1}) \\
v_{t}-(q_{r+1}-{\underline q}_{r+1}) \end{array}\right)=0 \ ,
\ee
where  $r$ is the index of the hierarchical flow and $p_{j}=p_{j}(n,t)$, $q_{j}=q_{j}(n,t)$ can be recursively generated from $p_{0}=1\ , q_{0}=0$ using
\bse\label{eq:com}\bea
p_{r+1}&=&\frac{1}{2}(q_{r}+\underline{q}_{r})+vp_{r} \ ,  \\
q_{r+1}&=&2u^2\sum_{l=0}^{r}p_{r-l}\ol{p}_{l}-\frac{1}{2}\sum_{l=0}^{r}q_{r-l}q_{l}\ .
\eea\ese
(This choice of notation can be found in \cite{BGHT}.)

The Toda equation is a widely recognized integrable differential-difference equation \cite{Flas1,Flas2}
\be\label{eq:fftoda} \left\{\begin{array}{ccc}
u_{t}& =& u(\ol{v}-v) \\
v_{t}& =& 2(u^2-\underline{u}^2)\end{array}\right.
\ee
where $u=u_{n}(t)=u(n,t)$, $v=v_{n}(t)=v(n,t)$ $u_{t}=\pp u(n,t)/\pp t$, $\ol{u}=u(n+1,t),\ \underline{u}=u(n-1,t)$,
$\ol{v}=v(n+1,t),\ \underline{v}=v(n-1,t)$
and can also be given by the corresponding Lax pair \cite{WT}:
\bse\bea\label{eq:fflax}
\lambda\psi &=&u\ol{\psi}+\underline{u}\underline{\psi} +v\psi  \\
\psi_{t} &=&u\ol{\psi}-\underline{u}\underline{\psi} \ .
\eea\ese
In both cases we see that the solutions depend on two independent variables $n,t$.  However in \cite{Joshi} we saw that it is
possible to reduce a differential-difference equation in two variables (\ref{eq:fftoda}) to a single variable in $\eta$
\bea\left\{\begin{array}{ccc}
-c_{0}H(\eta)+H_{\eta}& = &H(\eta)(G(\ol{\eta})-G(\eta)) \\
p_{0}-c_{0}G(\eta)+G_{\eta}& = & 2(H(\eta)^2-H(\underline{\eta})^2)
\end{array}\right.
\eea
where $p_{0},c_{0}$ are constants.  
This reduction led to an equation in which iterates and derivatives of the {\it same} variable appeared,
compared with (\ref{eq:fftoda}) which involves iterates in one variable and derivatives in another.

We focus on reductions to equations involving only one independent variable
\be
H_{\eta}=K(H,\ol{H},\underline{H}),\ H:\mathbb{R}\mapsto\mathbb{R}\label{eq:genH}
\ee
where $H_{\eta}=dH/d\eta$, $\ol{H}=H(\ol{\eta})$, $\underline{H}=H(\underline{\eta})$, with $\eta = \eta(n,t)$, $\ol{\eta}=\eta(n+1,t)$ and $\underline{\eta}=\eta(n-1,t)$.

\subsection{Direct Method} In 1989, Clarkson and Kruskal \cite{CK} used a ``direct approach'' to
find reductions of the Boussinesq equation and found new reductions which were not
captured by the classical Lie symmetry approach. This direct approach was later
shown to be related to ``non-classical'' symmetries of the Boussinesq equation.  However such a relationship has not yet been established for
differential-difference equations. In \cite{Joshi} a direct method was developed
to finding reductions of differential-difference equations.

The most general form for a reduction is
\be
u(n, t) = U(n, t,H(\eta),G(\eta)),\ v(n, t) = V (n, t,H(\eta),G(\eta))\ , \label{eq:genuv}
\ee
where $H$ and $G$ form a coupled system of equations of the form (\ref{eq:genH}). For Equations
(1.1), it turns out to be sufficient (see \S 3) to take the ansatz
\bse
\label{eq:ans}\bea
u(n,t)&=&a(n,t)+b(n,t)H(\eta) \\
v(n,t)&=&c(n,t)+d(n,t)G(\eta)
\eea
\ese
where $\eta=\eta(n,t)$.  Central to our argument are the following rules (stated for $a$, $b$
and $H$ for conciseness, but they apply also to $c$, $d$ and $G$)
\begin{itemize}
\item{Rule 1:} If $a(n,t)=a_{0}(n,t)+b(n,t)\Gamma(\eta)$, then we can take $\Gamma\equiv 0$ w.l.o.g. by substituting
$H(\eta)\mapsto H(\eta)-\Gamma(\eta)$.
\item{Rule 2:} If $b(n,t)$ has the form $b(n,t)=b_{0}(n,t)\Gamma(\eta)$, then we can take $\Gamma\equiv 1$ w.l.o.g. by substituting
$H(\eta)\mapsto H(\eta)/\Gamma(\eta)$.
\item{Rule 3:} If $\eta (n,t)$ is determined by an equation of the form $\Gamma= \eta_{0}(n,t)$, where $\Gamma$ is invertible, then we can take
$\Gamma(\eta)=\eta$ w.l.o.g. by substituting $\eta\mapsto\Gamma^{-1}(\eta)$.
\end{itemize}

{\bf Definition 1.1}.  Given non-zero, differentiable and invertible functions, $\Gamma(\eta)$, we refer to the transformations
$H(\eta)\rightarrow H(\eta)-\Gamma(\eta)$, $H(\eta)\rightarrow H(\eta)/G(\eta)$, and $\eta\rightarrow \Gamma^{-1}(\eta)$ as the reduction
transformations on (\ref{eq:ans}).

\subsection{Outline of Paper}  In Section 2, we will present direct reductions for the first two
flows in the hierarchy and then present an iterative scheme for the reduction of a general
$r^\textrm{th}$ flow.
In Section 3, we show that the ans\"atze (\ref{eq:ans}) in fact represents the general case and can be assumed without loss of generality.
In Section 4, we will present corresponding reductions of the Lax pairs for the hierarchy.
In Section 5 we will present our conclusions and remarks.

\section{Direct reductions of the Toda hierarchy}
\setcounter{equation}{0}
We perform reductions on the Toda hierarchy (\ref{eq:Toda}) considering the first two cases $r=1$ and $r=2$.
In the following we indicate generic functions of $\eta$ (which are assumed to be differentiable,
non-zero and invertible) by the notation $\Gamma_{j}(\eta)$.

\subsection{A direct reduction of the second flow of the Toda hierarchy}

\bpr
Using (\ref{eq:Toda}) we find that the second flow in the Toda hierarchy is
\be\label{eq:sf}
\textrm{TL}_{1}(u,v) =\left(\begin{array}{c} u_{t} = u(\ol{u}^2-\underline{u}^2+\ol{v}^2-v^2)   \\
v_{t} = 2(u^2(\ol{v}+v)-\underline{u}^2(v+\underline{v}))\end{array}\right)
\ee

Then given that the ans\"{a}tze (\ref{eq:ans}) holds, the only possible nonlinear
second-order reduction of (\ref{eq:sf}) of the form (\ref{eq:genuv}) that is unique up
to reduction transformations of $H$, $G$ and $\eta$ is given by
\bse\label{eq:rtsf}\bea
-c_{o}H+H_{\eta}&=&H(\ol{H}^2-\underline{H}^2+\ol{G}^2-G^2) \\
-c_{o}G+G_{\eta}&=&2(H^2(\ol{G}+G)-\underline{H}^2(G+\underline{G}))
\eea\ese
where the reduction is given by $\eta(n,t)= \nu(n)+\sg(t)\ ,\  \nu(n)$ being an arbitrary function, with
{\em\be
\sg(t) = \left\{\begin{array}{cc}
\frac{1}{c_{0}}\log(c_{0}t+c_{1})+c_{2}\ & \textrm{if}\ c_{0}\neq 0 \\
a_{0}t+a_{1}\ & \textrm{otherwise}\end{array}\right.
\ee}
where $c_{j}\ , j=0,1,2,\ a_{0},a_{1} $ are constants and the reductions of $u$ and $v$ are given by the following two respective cases
\begin{enumerate}
\item{Case $c_{0}\neq 0$:}
\bse\be
u(n,t) = \sqrt{\frac{1}{c_{0}t+c_{1}}}H(\eta)\ ,\ v(n,t) = \sqrt{\frac{1}{c_{0}t+c_{1}}}G(\eta)
\ee

\item{Case $c_{0}= 0$:}
\be
u(n,t) = \sqrt{a_{0}}H(\eta)\ , \ v(n,t) = \sqrt{a_{0}}G(\eta)
\ee\ese
\end{enumerate}

\epr

\begin{proof}
Under the ans\"{a}tze (\ref{eq:ans}), (\ref{eq:sf}) becomes
\bse\bea
a_{t}+b_{t}H+b\eta_{t} H_{\eta} & = & (a+bH)((\ol{a}+\ol{b}\ol{H})^2-(\underline{a}+\underline{b}\underline{H})^2+(\ol{c}+\ol{d}\ol{G})^2-(c+dG)^2) \nn \\
&&\label{eq:pqf} \\
c_{t}+d_{t}H+d\eta_{t} H_{\eta} & = & 2((a+bH)^2(\ol{c}+\ol{d}\ol{G}+c+dG)-(\underline{a}+\underline{b}\underline{H})^2(c+dG+\underline{c}+\underline{d}\underline{G}))\nn  \\
&&\label{eq:rsf} \eea\ese

We seek nonlinear reduced equations of the form (\ref{eq:genH}), therefore we require the terms $H\ol{H}^2,
H\underline{H}^2$, $H\ol{G}^2,HG^2$ to be present in the reduced equation (\ref{eq:pqf}) and we require the term $H^2 G$,
 to be present in the reduced equation (\ref{eq:rsf}).  Thus we require the following
\bse\bea
b\ol{b}^2&=&b\eta_{t}\Gamma_{1}(\eta) \label{eq:nl1} \\
b\underline{b}^2&=&b\eta_{t}\Gamma_{2}(\eta) \label{eq:nl2} \\
b\ol{d}^2&=&b\eta_{t}\Gamma_{3}(\eta) \label{eq:nl3} \\
bd^2&=&b\eta_{t}\Gamma_{4}(\eta) \label{eq:nl4} \\
db^2 &=&d\eta_{t}\Gamma_{5}(\eta) \label{eq:nl5}
\eea\ese
If we compare (\ref{eq:nl1}, \ref{eq:nl2}, \ref{eq:nl5}) we find that
$b=\sqrt{\underline{\eta}_{t}}\sqrt{\Gamma_{1}(\underline{\eta})}=\sqrt{\eta_{t}}\sqrt{\Gamma_{5}(\eta)}=\sqrt{\ol{\eta}_{t}}\sqrt{\Gamma_{2}(\ol{\eta})}$.
However by Rule 2, this implies that $\sqrt{\Gamma_{1}(\underline{\eta})}\equiv \sqrt{\Gamma_{5}(\eta)}\equiv \sqrt{\Gamma_{2}(\ol{\eta})}\equiv 1$
and therefore $\sqrt{\underline{\eta}_{t}}=\sqrt{\eta_{t}}=\sqrt{\ol{\eta}_{t}}$.
This implies that $\eta(n,t)=\nu(n)+\sg(t)$, where $\nu(n)$ is an arbitrary function and $\sg(t)$ is some differentiable function in $t$ only.

Furthermore by comparing (\ref{eq:nl3}, \ref{eq:nl4}) we find that
$d=\sqrt{\underline{\eta}_{t}}\sqrt{\Gamma_{3}(\underline{\eta})}\ $ $=\sqrt{\eta_{t}}\sqrt{\Gamma_{4}(\eta)}$.  However by Rule 2, this implies that
$\sqrt{\Gamma_{3}(\underline{\eta})}\equiv \sqrt{\Gamma_{4}(\eta)}\equiv 1$ and therefore
\bse\bea
b&=&\sqrt{\eta_{t}}=\sqrt{\sg_{t}}  \\
d&=&\sqrt{\eta_{t}}=\sqrt{\sg_{t}}=b
\eea\ese
In order to identify the remaining coefficients $a,c$ we must consider the other non-linear terms.  Thus by requiring that the terms
$H_{\eta}$ and $\ol{H}^2$ $HG$ all remain in the reduced equation, we require
\bse\bea
&& a\ol{b}^2=b\eta_{t}\Gamma_{6}(\eta)\ \Rightarrow\ a=b\Gamma_{6}(\eta)\ \Rightarrow a=0\quad \textrm{w.l.o.g.} \\
&& bcd=b\eta_{t}\Gamma_{7}(\eta)\ \Rightarrow\ c=d\Gamma_{7}(\eta)\ \Rightarrow\ c=0\quad \textrm{w.l.o.g.}
\eea\ese
by making use of the fact that $s=\sqrt{\eta_{t}}$ and of Rule 1.

Now if the linear term in $H$ on the left side of the equation remains in the reduced equation, then
\be
b_{t}=b\ol{b}^2\Gamma_{8}(\eta)\ \Rightarrow\ \sg_{tt} = 2(\sg_{t})^2\Gamma_{8}(\eta)\ .
\ee
However, since $\sg$ only depends on $t$, while $\eta$ also depends on $n$, this equations can only hold if $\Gamma_{8}$ is identically constant.
So we let this constant be $-c_{-1}$ and absorbing the value of 2 so that $-2c_{-1}=-c_{0}$, we find $\sg_{tt} = -(\sg_{t})^2c_{0}$, which we can integrate
with respect to $t$ giving
\be
\sg(t) = \left\{\begin{array}{cc}
\frac{1}{c_{0}}\log(c_{0}t+c_{1})+c_{2}\ & \textrm{if}\ c_{0}\neq 0 \\
a_{0}t+a_{1}\ & \textrm{if}\ c_{0}= 0.\end{array}\right.
\ee
where $c_{j}\ , j=0,1,2,\ a_{0},a_{1} $ are constants.

The reduced equations (\ref{eq:pqf}, \ref{eq:rsf}) are now
\bse\bea
-c_{o}H+H_{\eta}&=&H(\ol{H}^2-\underline{H}^2+\ol{G}^2-G^2)\nn \\
-c_{o}G+G_{\eta}&=&2(H^2(\ol{G}+G)-\underline{H}^2(G+\underline{G})) \nn
\eea\ese
\end{proof}

On comparison with the reduced equation for the Toda equation \cite{Joshi} (the first flow of the hierarchy), we see a number of
similarities, primarily the value of $\sg(t)$, which remains the same for the first two flows of the Toda hierarchy.  However for the
second flow, the reduced equations (\ref{eq:sf}) do not contain the additional constant term $p_{0}$ found in the first flow, thus we look at the
third flow to see whether this term reappears.

\subsection{A direct reduction of the third flow of the Toda hierarchy}

\bpr
Using (\ref{eq:Toda}) we find that the second flow in the Toda hierarchy is
{\small\bea\label{eq:tf}
\textrm{TL}_{2}(u,v) &=&\left(\begin{array}{c} u_{t} = u(\ol{u}^2(\ol{\ol{v}}+2\ol{v})+u(\ol{v}-v)-\underline{u}^2(2v+\underline{v})+\ol{v}^3-v^3)  \\
 v_{t} = 2(u^2(\ol{u}^2+u^2+\ol{v}^2+\ol{v}v+v^2)-\underline{u}^2(\underline{u}^2+\underline{\underline{u}}^2+v^2+v\underline{v}+\underline{v}^2))\end{array}\right) .  \nn \\
                         &&
\eea}

Then given that the ans\"{a}tze (\ref{eq:ans}) holds, the only possible nonlinear
second-order reduction of (\ref{eq:sf}) of the form (\ref{eq:genuv}) that is unique up
to reduction transformations of $H$, $G$ and $\eta$ is given by
\bse\bea
-c_{o}H+H_{\eta}&=&H(\ol{H}^2(\ol{\ol{G}}+2\ol{G})+H(\ol{G}-G)-\underline{H}^2(2G+\underline{G})+\ol{G}^3-G^3), \nn \\
&& \\
-c_{o}G+G_{\eta}&=&2(H^2(\ol{H}^2+H^2+\ol{G}^2+\ol{G}G+G^2)\nn  \\
&&-\underline{H}^2(\underline{H}^2+\underline{\underline{H}}^2+G^2+G\underline{G}+\underline{G}^2))
\eea\ese
where the reduction is given by $\eta(n,t)= \nu(n)+\sg(t)\ ,\  \nu(n)$ being an arbitrary function, with
{\em\be
\sg(t) = \left\{\begin{array}{cc}
\frac{1}{c_{0}}\log(c_{0}t+c_{1})+c_{2}\ & \textrm{if}\ c_{0}\neq 0 \\
a_{0}t+a_{1}\ & \textrm{otherwise}\end{array}\right.
\ee}
where $c_{j}\ , j=0,1,2,\ a_{0},a_{1} $ are constants and the reductions of $u$ and $v$ are given by the following two respective cases
\begin{enumerate}
\item{Case $c_{0}\neq 0$:}
\bse\be
u(n,t) = \sqrt[3]{\frac{1}{c_{0}t+c_{1}}}H(\eta)\ ,\ v(n,t) = \sqrt[3]{\frac{1}{c_{0}t+c_{1}}}G(\eta)
\ee

\item{Case $c_{0}= 0$:}
\be
u(n,t) = \sqrt[3]{a_{0}}H(\eta)\ , \ v(n,t) = \sqrt[3]{a_{0}}G(\eta)
\ee\ese
\end{enumerate}

\epr

\begin{proof}
Under the ans\"{a}tze (\ref{eq:ans}), (\ref{eq:tf}) becomes
\bse\bea
a_{t}+b_{t}H+b\eta_{t} H_{\eta} & = & (a+bH)[(\ol{a}+\ol{b}\ol{H})^2((\ol{\ol{c}}+\ol{\ol{d}}\ol{\ol{G}})+2(\ol{c}+\ol{d}\ol{G}))\nn  \\
&&+(a+bH)^2(\ol{c}+\ol{d}\ol{G}-c-dG)\nn \\
&&-(\underline{a}+\underline{b}\underline{H})^2(2(c+dG)+\underline{c}+\underline{d}\underline{G})+(\ol{c}+\ol{d}\ol{G})^3-(c+dG)^3]\nn  \\
&& \label{eq:pqf3} \\
c_{t}+d_{t}H+d\eta_{t} H_{\eta} & = & 2[(a+bH)^2((\ol{a}+\ol{b}\ol{H})^2+(a+bH)^2+(\ol{c}+\ol{d}\ol{G})^2\nn  \\
&&+(\ol{c}+\ol{d}\ol{G})(c+dG)+(c+dG)^2) \nn  \\
&&-(\underline{a}+\underline{b}\underline{H})^2((\underline{a}+\underline{b}\underline{H})^2+(\underline{\underline{a}}+\underline{\underline{b}}\underline{\underline{H}})^2+(c+dG)^2\nn  \\
&&+(c+dG)(\underline{c}+\underline{d}\underline{G})+(\underline{c}+\underline{d}\underline{G})^2)]\nn  \\
&&\label{eq:rsf3} \eea\ese
We seek nonlinear reduced equations of the form (\ref{eq:genH}), therefore we require the terms
 $H\ol{G}^3$, $HG^3$, $H\ol{H}^2\ol{G}$ to be present in the reduced equation (\ref{eq:pqf}) and we require the terms $H^2 G^2$,
to be present in the reduced equation (\ref{eq:rsf}).  Thus we require the following
\bse\bea
b\ol{d}^3&=&b\eta_{t}\Gamma_{1}(\eta) \label{eq:mnl1} \\
bd^3&=&b\eta_{t}\Gamma_{2}(\eta) \label{eq:mnl2} \\
b\ol{b}^2\ol{d}&=&b\eta_{t}\Gamma_{3}(\eta) \label{eq:mnl3} \\
b^2d^2&=&d\eta_{t}\Gamma_{4}(\eta) \label{eq:mnl4}
\eea\ese
If we compare (\ref{eq:mnl1}, \ref{eq:mnl2}) we find that
$d=\sqrt[3]{\underline{\eta}_{t}}\sqrt[3]{\Gamma_{1}(\underline{\eta})}=\sqrt[3]{\eta_{t}}\sqrt[3]{\Gamma_{2}(\eta)}$
However by Rule 2, this implies that $\sqrt[3]{\Gamma_{1}(\underline{\eta})}\equiv \sqrt[3]{\Gamma_{2}(\eta)} \equiv 1$
and therefore $\sqrt[3]{\underline{\eta}_{t}}=\sqrt[3]{\eta}$.
This implies that $\eta(n,t)=\nu(n)+\sg(t)$, where $\nu(n)$ is an arbitrary function and $\sg(t)$ is some differentiable function in $t$ only.
If we compare (\ref{eq:mnl3}, \ref{eq:mnl4}) we find that
$b=\sqrt[3]{\underline{\eta}_{t}}\sqrt[3]{\Gamma_{3}(\underline{\eta})}=\sqrt[3]{\eta_{t}}\sqrt[3]{\Gamma_{4}(\eta)}$
However by Rule 2, this implies that $\sqrt[3]{\Gamma_{3}(\underline{\eta})}\equiv \sqrt[3]{\Gamma_{4}(\eta)} \equiv 1$
and therefore
\bse\bea
b&=&\sqrt[3]{\eta_{t}}=\sqrt[3]{\sg_{t}}  \\
d&=&\sqrt[3]{\eta_{t}}=\sqrt[3]{\sg_{t}}=b
\eea\ese
To determine the remaining coefficients $a$ and $c$ we require the nonlinear terms $H\ol{H}^2$ and $\ol{H}^2\ol{G}$ to be present in the reduced
equation (\ref{eq:pqf}), hence we have the following
\bse\bea
b\ol{b}^2\ol{c}&=&b\eta_{t}\Gamma_{5}(\eta) \ \Rightarrow \ol{c}=d\Gamma_{5}(\eta)\ \Rightarrow \ol{c}=0  \\
a\ol{b}^2\ol{d}&=&b\eta_{t}\Gamma_{6}(\eta) \ \Rightarrow a=b\Gamma_{6}(\eta)\ \Rightarrow a=0
\eea\ese
by application of Rule 1.

Now if the linear term in $H$ on the left side of the equation remains in the reduced equation, then
\be
b_{t}=b\ol{b}^2\ol{d}\Gamma_{7}(\eta)\ \Rightarrow\ \sg_{tt} = 3(\sg_{t})^2\Gamma_{7}(\eta)\ .
\ee
However, since $\sg$ only depends on $t$, while $\eta$ also depends on $n$, this equations can only hold if $\Gamma_{7}$ is identically constant.
So we let this constant be $-c_{-1}$ and absorbing the value of 3 so that $-3c_{-1}=-c_{0}$, we find $\sg_{tt} = -(\sg_{t})^2c_{0}$, which we can integrate
with respect to $t$ giving
\be
\sg(t) = \left\{\begin{array}{cc}
\frac{1}{c_{0}}\log(c_{0}t+c_{1})+c_{2}\ & \textrm{if}\ c_{0}\neq 0 \\
a_{0}t+a_{1}\ & \textrm{if}\ c_{0}= 0.\end{array}\right.
\ee
where $c_{j}\ , j=0,1,2,\ a_{0},a_{1} $ are constants.

The reduced equations are now (\ref{eq:pqf}, \ref{eq:rsf}) are now
\bse\bea
-c_{o}H+H_{\eta}&=&H(\ol{H}^2(\ol{\ol{G}}+2\ol{G})+H(\ol{G}-G)-\underline{H}^2(2G+\underline{G})+\ol{G}^3-G^3),\nn \\
-c_{o}G+G_{\eta}&=&2(H^2(\ol{H}^2+H^2+\ol{G}^2+\ol{G}G+G^2)\nn  \\
&&-\underline{H}^2(\underline{H}^2+\underline{\underline{H}}^2+G^2+G\underline{G}+\underline{G}^2))
\eea\ese
\end{proof}

Having established a clear pattern from the reduced equations of the flows of the hierarchy, we can state the an expression
for the reductions of the $r^\textrm{th}$ flow of the hierarchy.

\subsection{A direct reduction of a general flow of the Toda hierarchy}

We use (\ref{eq:Toda}) and (\ref{eq:com}) and the results of \S 2.1 and \S 2.2 to give an iterative scheme. 
\bpr
The reduced $r^\textrm{th}$ flow of the Toda hierarchy can be generated by
\be\label{eq:nf}
\textrm{TL}_{r}(u,v) =\left(\begin{array}{c} -c_{0}H+H_{\eta}-H(p_{r+1}(\ol{\eta})-p_{r+1}(\eta))\\
 -c_{0}G+G_{\eta}+p_{0_{1}}-(q_{r+1}(\eta)-q_{r+1}(\underline{\eta}))\end{array}\right)=0 \ ,
\ee
where  $p_{j}=p_{j}(\eta)$, $q_{j}=q_{j}(\eta)$ can be recursively generated from $p_{0}=1\ , q_{0}=0$ using
\bse\label{eq:HGcom}\bea
p_{r+1}(\eta)&=&\frac{1}{2}(q_{r}(\eta)+q_{r}(\underline{\eta}))+Gp_{r}(\eta) \ ,  \\
q_{r+1}(\eta)&=&2H^2\sum_{l=0}^{r}p_{r-l}(\eta)p_{l}(\ol{\eta})-\frac{1}{2}\sum_{l=0}^{r}q_{r-l}(\eta)q_{l}(\eta)\ .
\eea\ese
and the value $p_{0_{1}}$ represents the additional term in the first flow \cite{Joshi}.
The reduction is given by $\eta(n,t)= \nu(n)+\sg(t)\ ,\  \nu(n)$ being an arbitrary function, with
{\em\be
\sg(t) = \left\{\begin{array}{cc}
\frac{1}{c_{0}}\log(c_{0}(t)+c_{1})+c_{2}\ & \textrm{if}\ c_{0}\neq 0 \\
a_{0}t+a_{1}\ & \textrm{otherwise}\end{array}\right.
\ee}
where $c_{j}\ , j=0,1,2,\ a_{0},a_{1} $ are constants and the reductions of $u$ and $v$ are given by the following two respective cases
\begin{enumerate}
\item{Case $c_{0}\neq 0$:}
\bse\be
u(n,t) = \sqrt[r]{\frac{1}{c_{0}t+c_{1}}}H(\eta)\ ,\ v(n,t) = \sqrt[r]{\frac{1}{c_{0}t+c_{1}}}G(\eta)
\ee

\item{Case $c_{0}= 0$:}
\be
u(n,t) = \sqrt[r]{a_{0}}H(\eta)\ , \ v(n,t) = \sqrt[r]{a_{0}}G(\eta)
\ee\ese
\end{enumerate}
\epr

\begin{remark}Within the hierarchy each set of reduced equations (\ref{eq:nf}) form a system of differential-difference equations, which evolves on a sequence of domains
containing points
\be
\mathcal{P}=\{\eta_{0},\ol{\eta_{o}},\ol{\ol{\eta_{0}}},\ldots\}\ ,
\ee
where if $\eta_{0}=\nu(n)+\sg(t)$, then $\ol{\eta_{0}}=\nu(n+1)+\sg(t)$.  In any interior of any domain in $n$, where the mapping $\nu(n)\rightarrow \nu(n+1)$ is
defined, we get a semi-infinite chain of points $\mathcal{P}$ and a corresponding sequence of domains on which these iterates are defined.  Since $\nu(n)$ is an
arbitrary function we have an infinite-dimensional family of reductions.
\end{remark}
\section{Generalization of the ans\"atze}
\setcounter{equation}{0}
In \cite{Joshi} it was shown that the ans\"atze (\ref{eq:ans}) represent the general case for the first flow in the Toda hierarchy, the Toda equation.
For the other flows within the hierarchy we show how the ans\"atze represent the general case (\ref{eq:ans}) for the second flow.  It can be shown that this holds true for
the $r^\textrm{th}$ flow of the hierarchy.  Consider the general reduction
\bse\bea
u(n, t) & =& U(n, t,H(\eta(n,t)),G(\eta(n,t))) \\
v(n, t) &=& V(n, t,H(\eta(n,t)),G(\eta(n,t)))\ .
\eea\ese
Under these transformations the second flow of the Toda equation (\ref{eq:sf}) becomes
\bse\label{eq:gsf}\bea
 U_{t}+U_{H}H_{\eta}\eta_{t}+U_{G}G_{\eta}\eta_{t} & = &  U(\ol{U}^2-\underline{U}^2+\ol{V}^2-V^2) \ ,  \\
 V_{t}+V_{H}H_{\eta}\eta_{t}+V_{G}G_{\eta}\eta_{t} & = & 2(U^2(\ol{V}+V)-\underline{U}^2(V+\underline{V}))\ .
\eea\ese
For the reduced equations to each contain the nonlinear terms in $H,G$ we require
\be
\ol{U}^2=\underline{U}^2\Gamma_{1}(\eta,H,G)\quad \ol{V}=V\Gamma_{2}(\eta,H,G) \ ,
\ee
then rewriting $\Gamma_{1}$ and $\Gamma_{2}$ in these equations appropriately, we can sum up to get
\be
U=e(t)\Gamma_{3}(\eta, H, G)+f(t)\quad V=j(t)\Gamma_{4}(\eta,H,G)\ .
\ee
Redefining $\Gamma_{3}=\wt{H}$ and $\Gamma_{4}=\wt{G}$ where $\wt{H}$ and $\wt{G}$ are new variables replacing $H$ and $G$, we regain the linear ans\"atze
assumed earlier (\ref{eq:ans}).
It is possible to show this for the other flows in the hierarchy.

\section{Lax Pairs}
\setcounter{equation}{0}

The general Lax pair \cite{BGHT} for the Toda hierarchy is given by
\bse\label{eq:glax}\bea
\lambda\psi &=&u\ol{\psi}+\underline{u}\underline{\psi} +v\psi \label{eq:llax} \\
\psi_{t} &=&2uP_{r}\ol{\psi}-Q_{r+1}\psi \label{eq:dlax}
\eea\ese
where
$P_{r}=P_{r}(\lambda,n,t)\ , Q_{r}=Q_{r}(\lambda,n,t)$ are monic
polynomials in the spectral parameter $\lambda$ of the type
\bse\label{eq:pqlax}\bea
P_{r} & = & \sum_{j=0}^{r}\lambda^{j}p_{r-j} \ , \\
Q_{r} & = &\lambda^{r+1}+\sum_{j=0}^{r}\lambda^{j}q_{r-j}-p_{r+1} .
\eea\ese
The Lax pair for the hierarchy can be derived independently of this formula, by looking at the compatibility between (\ref{eq:llax}) and
(\ref{eq:dlax}), setting $P_{r}=\sum_{j=0}^{r-1}\ar_{j}\lambda^{j}$, $Q_{r}=\sum_{j=0}^{r}\bb_{j}\lambda^{j}$ and finding appropriate values
for $\ar_{j},\bb_{j}$.

Then using (\ref{eq:glax}), the Lax pair for the Toda
equation \cite{WT} is given as \bse \bea
&& u\ol{\psi}+\underline{u}\underline{\psi}+v\psi=\lambda\psi \ , \\
&& \psi_{t}=2uP_{0}\ol{\psi}-Q_{0}\psi\ , \eea \ese
where $P_{0}=1\ ,\ Q_{0}=(\lambda-v)$.
To illustrate this notation we present the reduction for the second and third flows, then give the reduction
for the general Lax pair.

\subsection{Second Flow}

Using (\ref{eq:glax}) and (\ref{eq:pqlax}) for $r=2$ we derive the Lax pair for the second flow of the hierarchy
\bse\label{eq:sflax}\bea
\lambda\psi &=&u\ol{\psi}+\underline{u}\underline{\psi} +v\psi  \\
\psi_{t} &=&2u(\lambda +v)\ol{\psi}-(\lambda^2 +u^2-\underline{u}^2-v^2)\psi
\eea\ese
then from the results of Proposition 2.1, we find from (\ref{eq:sflax})
\be
\pm\sqrt{\sg '(t)}H\ol{\psi}\pm\sqrt{\sg '(t)}\underline{H}\underline{\psi}\pm\sqrt{\sg '(t)}G\psi=\lambda\psi\  .
\ee
We define a new spectral parameter $\zeta=\frac{\lambda}{\sqrt{\sg '}}$ and choose a definite sign in $b=\pm\sqrt{\sg '}$, say the positive sign,
then we obtain for $\phi(\eta ,\zeta)=\psi(n,t)$
\be
\psi_{t}=\zeta_{t}\phi_{\zeta}+\sg '\phi_{\eta}=\sg ' (c_{0}\zeta\phi_{\zeta}+\phi_{\eta})\ .
\ee
Thus we get the reduced second flow Lax pair
\bse\label{eq:rrsf}\bea
\zeta\phi &=&H\ol{\phi}+\underline{H}\underline{\phi} +G\phi \label{eq:rsfa} \\
c_{0}\zeta\phi_{\zeta}+\phi_{\eta} &=&2H(\zeta +G)\ol{\phi}-(\zeta^2 +H^2-\underline{H}^2-G^2)\phi \label{eq:rsfb}
\eea\ese

We note that the character of the Lax pair has changed from a spectral problem to a monodromy
problem, since now derivatives in $\zeta$ also appear in the linear problem. By differentiating
equation (\ref{eq:rsfa}) in two different ways, once with respect to $\zeta$ and once with respect to $\eta$, and
using (\ref{eq:rsfb}) to replace $\zeta$ , while using (\ref{eq:rsfa}) to replace $\ol{\phi}$, we can show that the compatibility
conditions for the linear system (\ref{eq:rrsf}) are precisely equations (\ref{eq:rtsf}).

\subsection{Third Flow}

Using (\ref{eq:glax}) and (\ref{eq:pqlax}) for $r=3$ we derive the Lax pair for the second flow of the hierarchy
\bse\label{eq:tflax}\bea
\lambda\psi &=&u\ol{\psi}+\underline{u}\underline{\psi} +v\psi  \\
\psi_{t} &=&2u(\lambda^2+\lambda v+u^2+\underline{u}^2+v^2)\ol{\psi} \nn  \\
&&-\left(\lambda^3+\lambda 2u^2+u^2\ol{v}-\underline{u}^2(2v+\underline{v})-v^3\right)\psi
\eea\ese
then from the results of Proposition 2.2, we find from (\ref{eq:tflax})
\be
\sqrt[3]{\sg '(t)}H\ol{\psi}+\sqrt[3]{\sg '(t)}\underline{H}\underline{\psi}+\sqrt[3]{\sg '(t)}G\psi=\lambda\psi\  .
\ee
Using the new spectral parameter $\zeta$, and the change of variables $\phi(\eta ,\zeta)=\psi(n,t)$, we get the reduced third flow Lax pair
\bse\bea
\zeta\phi &=&H\ol{\phi}+\underline{H}\underline{\phi} +G\phi \\
c_{0}\zeta\phi_{\zeta}+\phi_{\eta} &=&2H(\zeta^2+\zeta G+H^2+\underline{H}^2+G^2)\ol{\phi} \nn  \\
&&-\left(\zeta^3++\zeta 2H^2 +H^2\ol{G}-\underline{H}^2(2G+\underline{G})-G^3\right)\phi\ . \nn  \\
&&
\eea\ese

Similarly for the third flow we find we again have a monodromy problem,
though we can consider the compatibility in a similar manner to the second flow.


\begin{remark}It is clear how the reduced hierarchy for the Toda equation progresses and thus we simply state the general
reduced Lax pair to calculate up to the $r^\textrm{th}$ flow:
\bse\bea
\zeta\phi &=&H\ol{\phi}+\underline{H}\underline{\phi} +G\phi \\
c_{0}\zeta\phi_{\zeta}+\phi_{\eta} &=&2HP_{r}\ol{\phi}-Q_{r+1}\phi
\eea\ese
where
$P_{r}=P_{r}(\zeta,\eta)\ , G_{r}=G_{r}(\zeta,\eta)$ are monic
polynomials in the spectral parameter $\zeta$ of the type
\bse\bea
P_{r}(\zeta,\eta) & = & \sum_{j=0}^{r}\zeta^{j}p_{r-j}(\eta) \ , \\
Q_{r}(\zeta,\eta) & = &\zeta^{r+1}+\sum_{j=0}^{r}\zeta^{j}q_{r-j}(\eta)-p_{r+1}(\eta)
\eea\ese
where $p(\eta),q(\eta)$ are given by (\ref{eq:HGcom}).
\end{remark}

\section{Conclusions and Remarks}

In Section 2, we have extended the reduction of the Toda equation (given in \cite{Joshi}), to the Toda hierarchy and have thus
obtained the most general possible reductions for the ans\"{a}tze (\ref{eq:ans}).  Our results from the second and third flows led to the
iterative scheme for a reduction up to the $r^{\textrm{th}}$ flow.  In Section 3, we showed that the ans\"{a}tze assumed in fact represents the most general case.
Finally in Section 4, we showed that the reduced equations from the hierarchy also inherit a linear problem.


The inheritance of a linear problem through the reduction indicates that the set of equations
(\ref{eq:nf}) are integrable. These results suggest that the system (\ref{eq:nf}) is analogous
to the well known reductions of completely integrable partial differential equations,
namely the classical Painlev\'e equations. However, questions remain open on how
close such an analogy might be.

\bibliographystyle{unsrt}
\bibliography{toda}
\end{document}